\documentclass[aps,prb,epsf,twocolumn,amsmath,amssymb,showpacs,floatfix,superscriptaddress]{revtex4-1}

\usepackage{amsmath}
\usepackage{amssymb}
\usepackage{graphicx}
\usepackage{xspace}		
\usepackage{multirow}	

\usepackage{dcolumn}		
\usepackage{bm}			
\usepackage{subfigure}	
\usepackage{hyperref}	
  \hypersetup{colorlinks = true, 	
  citecolor = blue,		
  linkcolor = red, 		
  breaklinks = true,
  backref = true			
  }

\usepackage{upgreek}

\begin{document}
\title{Pt$_3$Zr(0001): A substrate for growing well-ordered ultrathin zirconia films by oxidation}

\author{Moritz Antlanger}
\affiliation{Institute of Applied Physics, Vienna University of Technology, 1040 Vienna, Austria}

\author{Wernfried Mayr-Schm{\"o}lzer}
\affiliation{Institute of Applied Physics, Vienna University of Technology, 1040 Vienna, Austria}
\affiliation{Center for Computational Materials Science, Vienna University of Technology, 1040 Vienna, Austria}

\author{Ji\v{r}\'{\i} Pavelec}
\affiliation{Institute of Applied Physics, Vienna University of Technology, 1040 Vienna, Austria}

\author{Florian Mittendorfer}
\affiliation{Institute of Applied Physics, Vienna University of Technology, 1040 Vienna, Austria}
\affiliation{Center for Computational Materials Science, Vienna University of Technology, 1040 Vienna, Austria}

\author{Josef Redinger}
\affiliation{Institute of Applied Physics, Vienna University of Technology, 1040 Vienna, Austria}
\affiliation{Center for Computational Materials Science, Vienna University of Technology, 1040 Vienna, Austria}

\author{Peter Varga}
\affiliation{Institute of Applied Physics, Vienna University of Technology, 1040 Vienna, Austria}

\author{Ulrike Diebold}
\affiliation{Institute of Applied Physics, Vienna University of Technology, 1040 Vienna, Austria}

\author{Michael Schmid}
\affiliation{Institute of Applied Physics, Vienna University of Technology, 1040 Vienna, Austria}

\date{\today}


\begin{abstract}

We have studied the surface of pure and oxidized Pt$_3$Zr(0001) by scanning tunneling microscopy (STM), Auger electron microscopy, and density functional theory (DFT). The well-annealed alloy surface shows perfect long-range chemical order. Occasional domain boundaries are probably caused by nonstoichiometry.  Pt$_3$Zr exhibits ABAC stacking along [0001]; only the A-terminated surfaces are seen by STM, in agreement with DFT results showing a lower surface energy for the A termination. DFT further predicts a stronger inward relaxation of the surface Zr than for Pt, in spite of the larger atomic size of Zr.

A closed ZrO$_2$ film is obtained by oxidation in $10^{-7}$ mbar O$_2$ at 400\,$^\circ$C and post-annealing at $\approx 800\,^\circ$C. The oxide consists of an O-Zr-O trilayer, equivalent to a (111) trilayer of the fluorite structure of cubic ZrO$_2$, but contracted laterally. The oxide forms a $(\sqrt{19} \times \sqrt{19})$R23$^\circ$ superstructure. The first monolayer of the substrate consists of Pt and contracts, similar to the metastable reconstruction of pure Pt(111).  DFT calculations show that the oxide trilayer binds rather weakly to the substrate. In spite of the O-terminated oxide, bonding to the substrate mainly occurs via the Zr atoms in the oxide, which strongly buckle down towards the Pt substrate atoms, if near a Pt position. According to DFT, the oxide has a bandgap; STM indicates that the conduction band minimum lies $\approx 2.3$ eV above $E_\mathrm{F}$.

\end{abstract}


%

\maketitle


\section{Introduction}

Zirconia (ZrO$_2$) in pure and doped form is an important material in various fields, with applications in engineering (zirconia toughened alumina, ZTA; zirconia-based thermal barrier coatings for gas turbines\cite{Aygun}), ceramics for dentistry (tetragonal zirconia polycrystal, TZP),\cite{Denry} and as a high-$k$ dielectric in microelectronics.\cite{BauerHiK}  Zirconia is also a catalyst support and a catalyst on its own.\cite{Tanabe,Steiner}  With its wide bandgap ($> 5$ eV), zirconia is a perfect insulator. It remains an electronic insulator, even when it becomes an ionic conductor at high temperatures, or when doped with trivalent metals (Y, Sc) introducing oxygen vacancies.  This property forms the basis of its applications in solid-state electrochemistry, where it is an electrolyte  in solid-oxide fuel cells\cite{Steele} and gas sensors.\cite{Goepel}  The low electric conductivity of zirconia is not only a blessing but also a curse:  While the surface properties of ZrO$_2$ are crucial for many of its applications, detailed studies of ZrO$_2$ surfaces are hampered by the fact that electron-based methods like scanning tunneling microscopy (STM) cannot be applied to this material.

To overcome this problem, ultrathin ZrO$_2$ films have been prepared by deposition and oxidation of Zr on a Pt(111) substrate, resulting in (111)-oriented films with cubic fluorite ZrO$_2$ structure \cite{Maurice, Meinel03, Meinel04}.  On these films, STM images have been successfully acquired, but it was difficult to obtain atomic resolution.\cite{Meinel03, Meinel04, Meinel06}  Growing ultrathin films this way is not straightforward because it is  difficult to evaporate zirconium: It has a high melting point (too high for evaporation from a heated crucible) and low vapor pressure at the melting point.  Thus, evaporation from a rod in standard ultrahigh-vacuum (UHV) electron-beam evaporators is a lengthy process, prone to contamination with impurities from the residual gas.

Another route towards synthesis of ultrathin oxide films is via the oxidation of a suitable metal alloy.  Experience from ultrathin alumina films shows that such an alloy should fulfill two criteria:  It should be be based on a metal that is rather inert towards to oxidation, and the alloy should only reluctantly liberate the reactive metal, so that it will be oxidized rather slowly, an important prerequisite for the formation of well-ordered oxide films.  This is the case for alloys with a low concentration of the reactive metal \cite{NapteschnigCuAl}, and so-called superalloys, which show strong chemical ordering and high stability, impeding diffusion of the reactive metal from the bulk to the surface [Examples include the well-known alumina films on NiAl(110)\cite{LibudaNiAlOx, KresseScience} and Ni$_3$Al(111)\cite{Ni3Al}]. In the present work, we use the superalloy Pt$_3$Zr(0001) as such a substrate for the growth of ultrathin ZrO$_2$ films.

\begin{figure}[tb]
  \centering
    {\includegraphics [width=5.5cm]
    {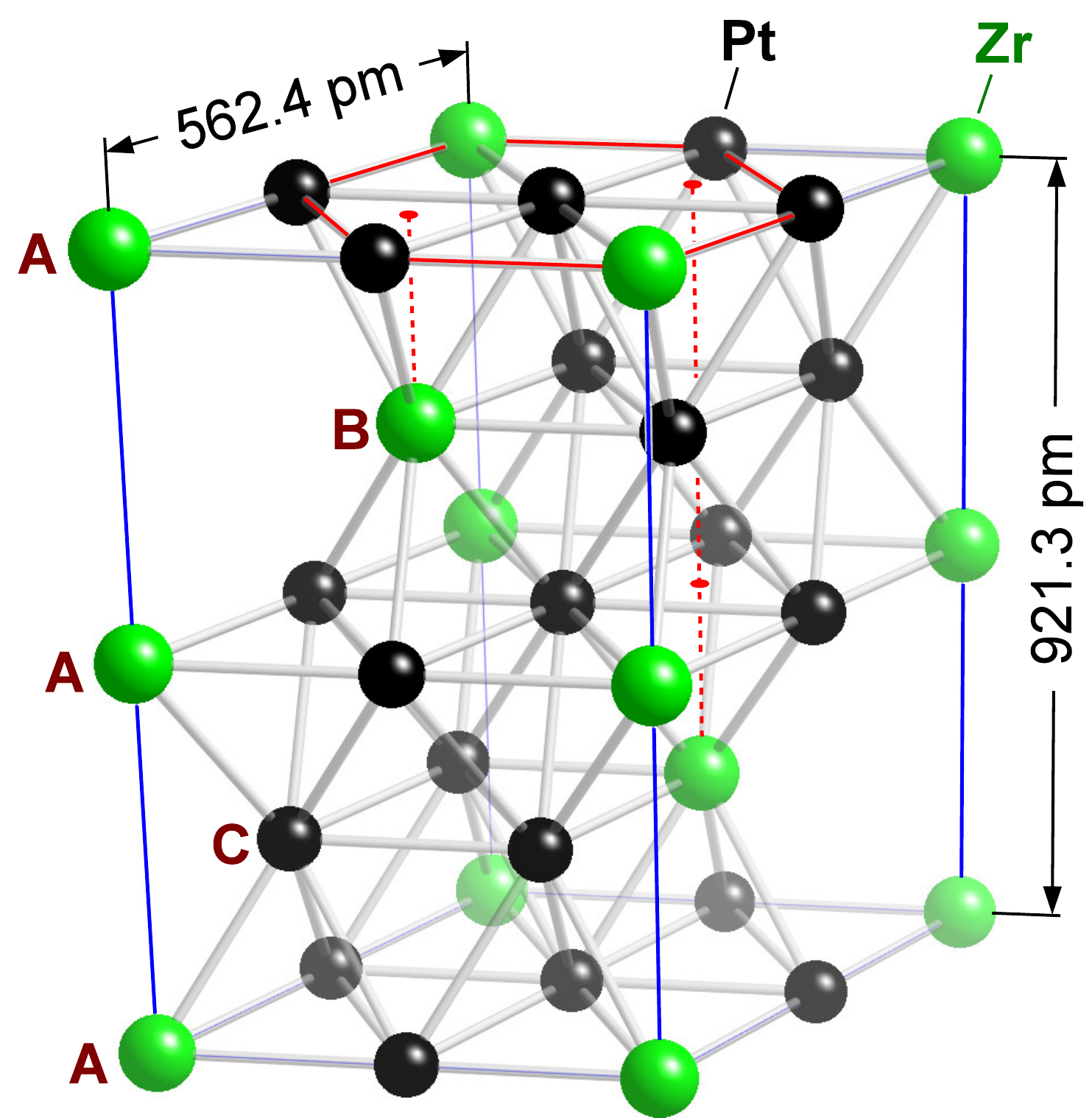} }
    \caption{(Color online) Schematic view of the Pt$_3$Zr unit cell (D0$_{24}$ structure).  Dotted (red) vertical lines indicate the stacking of Zr atoms in the B and C layers. Lattice constants from Ref.\ \onlinecite{PtZrLB}.}
  \label{FigNi3Ti}
  \end{figure}

Pt$_3$Zr and other AB$_3$ alloys with elements from groups 4 and 9--10 are also called ``Engel-Brewer'' alloys. It was noted that the average number of $d$ electrons per atom in such alloys is $\approx 5$, and such a half-filled $d$ shell guarantees exceptionally strong bonding.\cite{Brewer, HumeRothery}  A more ``chemical'' interpretation for the high stability would consider electron transfer from the electropositive Zr to Pt, emptying the $d$ shell of Zr and completely filling that of Pt.  Pt$_3$Zr has a high formation energy of $-128$ kJ/g\,atom (Ref.\ \onlinecite{Srikrishnan}) and a melting point of 2427\,K, significantly higher than Pt (2042\,K) and Zr (2128\,K).\cite{PtZrLB}   Pt$_3$Zr crystallizes in the Ni$_3$Ti (D0$_{24}$)  structure, shown schematically in Fig.\ \ref{FigNi3Ti}.  This  structure exhibits ABAC stacking; it can be considered a hybrid of the face-centered cubic and hexagonally close-packed structures.  With each Zr atom surrounded by Pt neighbors only, the number of bonds between unequal atoms is maximized.  This chemical order prevents easy gliding of dislocations, making Ni$_3$Ti-type materials superalloys, whose strength can even increase with temperature (yield strength anomaly).\cite{Hagihara}

We describe the (0001) surface of this alloy in section \ref{secAlloy} and the preparation and properties of the ultrathin ZrO$_2$ film on this alloy in section \ref{secOxide}. The experimental work is supplemented by a density-functional theory (DFT) study of the alloy (section \ref{secAlloyDFT}) and the ultrathin oxide (section \ref{secOxideDFT}).

\section{Experimental and Computational Methods}

The experiments were performed in a two-chamber ultrahigh-vacuum (UHV) system, with a preparation chamber for sputtering, annealing and oxidation, and an analysis chamber for STM and Auger electron spectroscopy (AES). Both chambers had a base pressure below 10$^{-10}$\,mbar. For cleaning we used 2\,keV Ar$^+$ ions with a current density between 5 and 10\,$\mathrm{\mu}$A\,cm$^{-2}$ and typical sputter times of 20 minutes. An electron-beam heater was used for annealing.  The thermocouple was attached to the sample holder and not to the single crystal, thus, the thermocouple does not show the correct sample temperature at high temperatures. Temperatures above 800\,$^\circ$C were therefore measured with a disappearing-filament pyrometer. For oxidation, we used a directional doser with a microcapillary plate in close proximity to the sample; the pressure at the sample position is estimated as $10^{-7}$\,mbar at a chamber pressure of $5 \times 10^{-9}$\,mbar.

The STM images were acquired in constant current mode at room temperature with electrochemically etched W tips; positive voltages refer to tunneling into unoccupied states of the surface (positive sample).  We also studied the Fourier transforms of STM images. When both the substrate and the overlayer lattice can be seen in the Fourier transform, it contains essentially the same information as a LEED pattern, except for the spot intensities.  Using the Fourier transformed STM image has the advantage that we can select only one domain, while LEED patterns of the oxide are more complex due to the superposition of the mirror domain.  For an accurate determination of lattice constants in STM images, we followed the procedure in Ref.\ \onlinecite{deSantisCoO}, based on selected Fourier components transformed back to real space; comparison of the substrate and overlayer lattice in the same image results in an error well below 1\%.  A cylindrical mirror analyzer with a concentric electron source (3\,keV) was used for AES.

The DFT studies were performed with the Vienna Ab-initio Simulation Package (VASP), using the projected augmented wave (PAW) method  \cite{bloechl1994,kressejoubert1999}  and an energy cutoff of 400 eV. The generalized gradient approximation (GGA) proposed by Perdew, 
Burke and Ernzerhof\cite{Perdew1996} (PBE) was used to treat the electronic exchange and correlation effects.   
A $\Gamma$-centered  $12 \times 12 \times 12$ Monkhorst-Pack mesh \cite{monkhorstpack1976} was used for the calculation of the Pt$_3$Zr bulk properties,  and a $12 \times 12 \times 1$ k-point mesh for the surface calculations. All structures have been relaxed until the residual forces were $< 0.01$\,eV/\AA.  As we find only a weak interaction between the Pt$_3$Zr surface and the ZrO$_2$ layer, we also investigated the influence of non-local effects within the framework of van-der-Waals DFT (vdW-DF), using the optB88 functional. \cite{klimes2011}
\section{The P\lowercase{t}$_3$Z\lowercase{r}(0001) surface}\label{secAlloy}
\subsection{STM results}\label{secAlloySTM}

\begin{figure}[tb]
  \centering
    {\includegraphics [width=8cm]
    {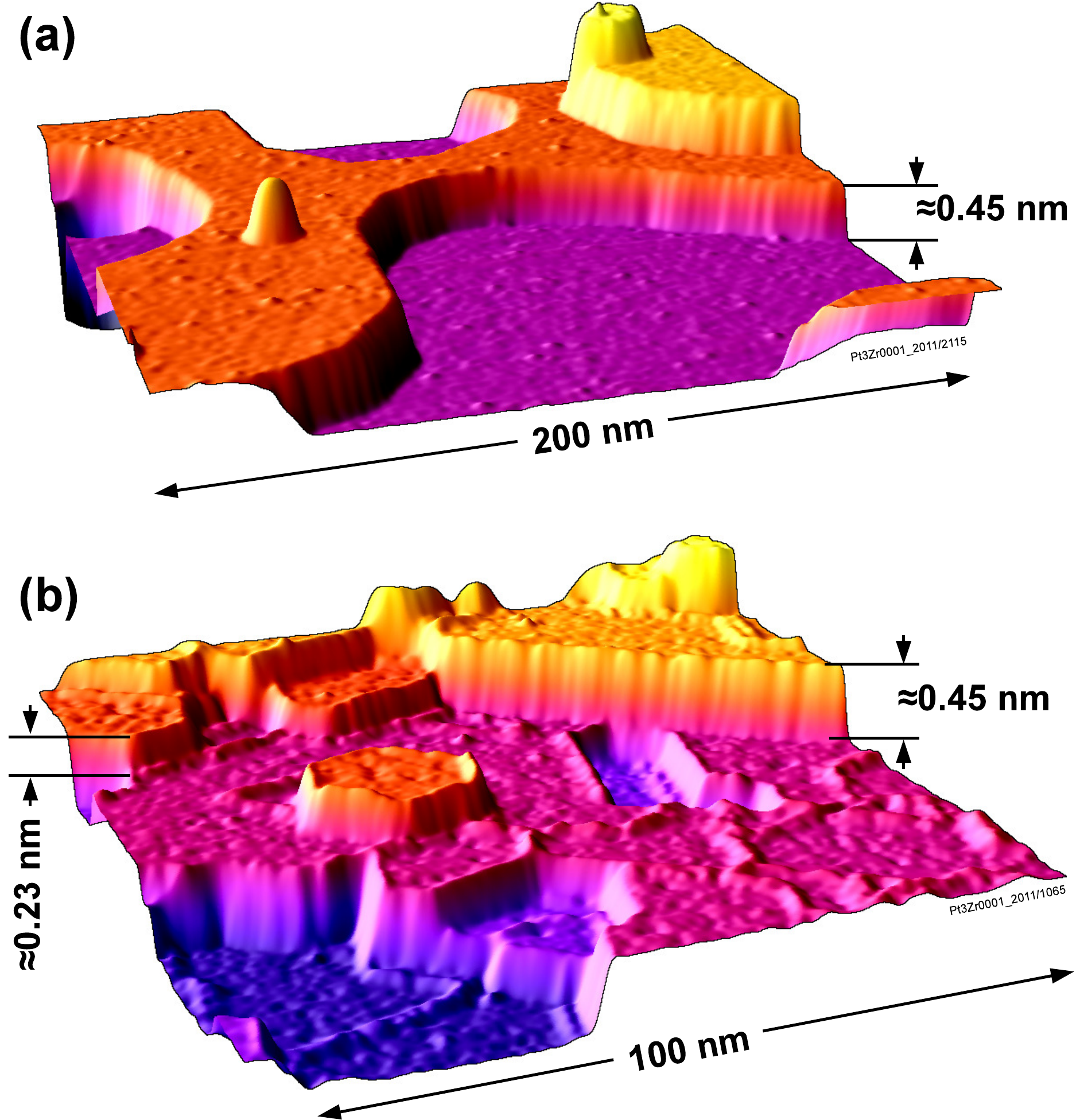} } 
    \caption{(Color online) STM images of the Pt$_3$Zr surface annealed at 900\,$^\circ$C; note the different image scales. (a) Annealing after 20\,min sputtering while cooling the sample from 400 to 100\,$^\circ$C.
    (b) Annealing after 30\,min sputtering of the cold sample. }
  \label{FigPt3ZrSteps}
  \end{figure}

After a few cycles of sputtering and annealing, the Pt$_3$Zr(0001) surface showed large terraces, and AES showed no impurities except for a small concentration of residual oxygen [few \% of a monolayer (ML)].  On a well-prepared sample, all steps have a height of $\approx 0.45$\, nm [Fig.\ \ref{FigPt3ZrSteps}(a)]. This indicates that the chemical order is perfect over distances $>$\, 100\,nm, and that only one type of surface termination is present, either A or B/C (cf. Fig.\ \ref{FigNi3Ti}).  The annealing temperature required for such a surface is $\approx 900\,^\circ$C or higher, depending on the history of the sample (see below).

\begin{figure}[tb]
  \centering
    {\includegraphics [width=8.5cm]
    {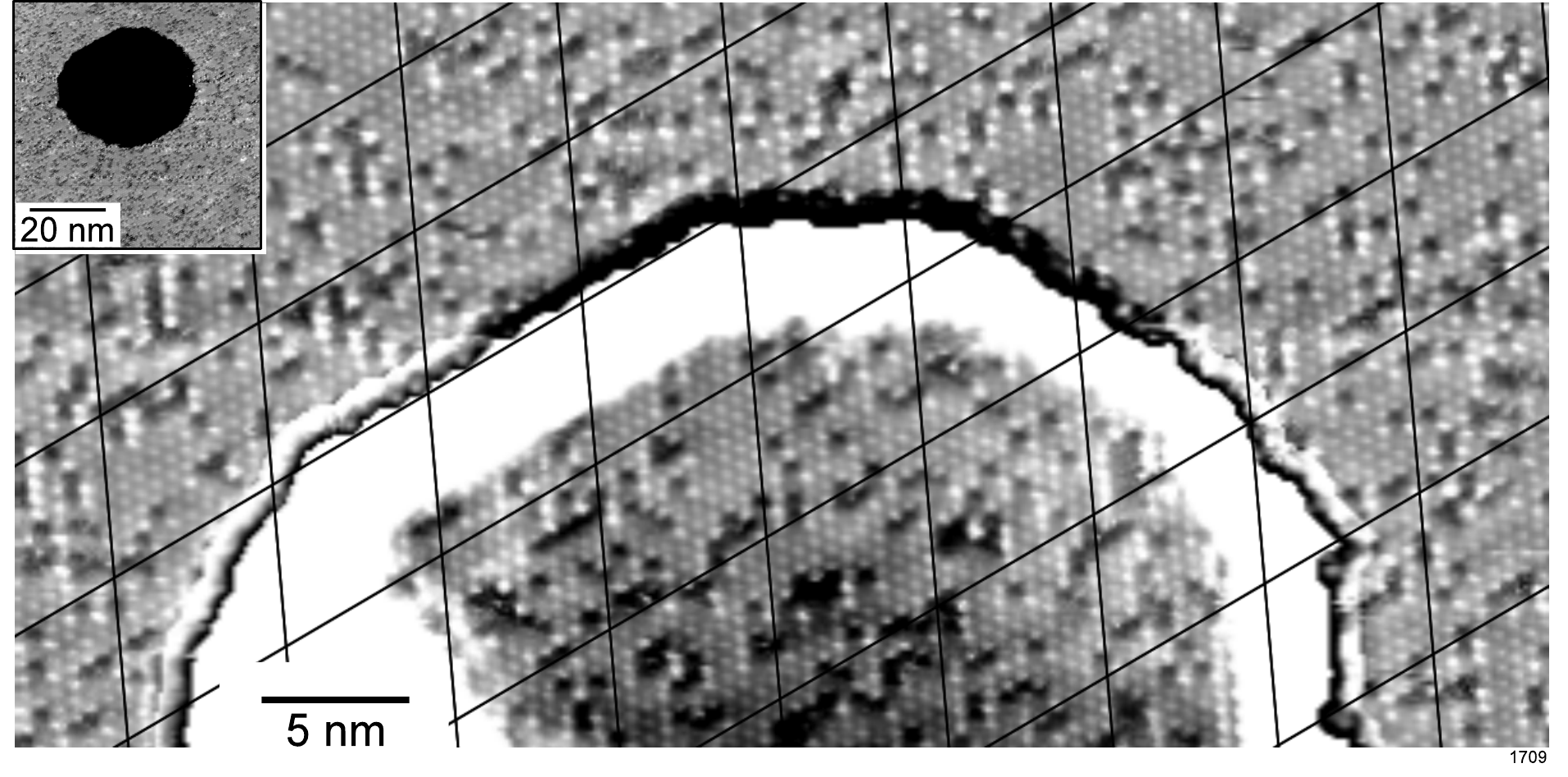} }
    \caption{STM image of a 0.45\,nm deep hole on the annealed Pt$_3$Zr surface. The inset shows an overview image. The brightness of the terraces is adjusted to show both layers simultaneously, and a slight nonlinear distortion of the image due to creep of the piezo scanner was removed (assuming that the top-layer lattice should be undistorted).  Two sets of parallel lines show that the lattices of bright (Zr) atoms in both layers are in registry. }
  \label{FigPt3ZrHole}
  \end{figure}

STM images with sufficiently high resolution show a hexagonal array of protrusions with a lattice constant of 0.56\,nm, i.e., one protrusion per unit cell of the alloy (Fig.\ \ref{FigPt3ZrHole}).  Alloys with strong chemical ordering commonly show a bulk-terminated surface.\cite{Varga}  Comparison with Fig.\ \ref{FigNi3Ti} strongly suggests that the protrusions are Zr, the element with the larger atomic radius (0.16 vs.\ 0.14\,nm for Zr and Pt, respectively).  So far, we could not determine whether the missing protrusions in the STM images are antisite defects (Zr replaced by Pt) or Zr atoms appearing dark due do oxygen impurities.

As the step heights of well-annealed defect-free samples are always 0.45\,nm or an integer multiple thereof [Fig.\ \ref{FigPt3ZrSteps}(a)], the alloy termination layer is either always an A layer or never an A layer (apart from a 180$^\circ$ rotation, the B and C layers are equivalent; see Fig.\ \ref{FigNi3Ti}).  STM images with clearly resolved 0.56\,nm alloy lattice on adjacent layers allow us to determine the registry of the surface lattice in these layers.  However, it is difficult to extrapolate the lattice across steps, because of drift and image distortions.  In case of a hole surrounded by a well-resolved atomic lattice, we can determine the continuation of the surface lattice more easily.   Fig.\ \ref{FigPt3ZrHole} shows that the lattices of Zr atoms in the top layer and at the bottom of the hole are in registry.  In the Pt$_3$Zr structure, all A layers have their Zr atoms above each other (Fig.\ \ref{FigNi3Ti}); therefore we can conclude that the surface layers are A layers.

Sample preparation can also yield a higher density of steps, some of which are also $\approx 0.23$\,nm high [Fig.\ \ref{FigPt3ZrSteps}(b)].  In addition, we find a network of lines, parallel to the close-packed directions of the metal, i.e., $\langle2\bar{1}\bar{1}0\rangle$. Some lines appear as protrusions (ridges) with a height of $\approx 0.05$\,nm, others as depressions with a similar depth.  These lines are not observed on well-prepared surfaces, which only exhibit 0.45\,nm steps.  These ridges are often found at steps, then they pin the steps, and the steps are straight, as opposed to well-prepared samples, where steps can be curved and have any direction.  We interpret these lines as domain boundaries of the alloy intersecting the surface.

While heating to high temperature helps reducing the number of such domain boundaries, we found that preparation with an annealing temperature of $\approx$\,900\,$^\circ$C can lead to both, well-prepared surfaces and surfaces with a large number of steps and defects (Fig.\ \ref{FigPt3ZrSteps}).  We could reproducibly avoid the domain boundaries by sputtering not at room temperature (RT) but rather soon after high-temperature annealing, while the sample was cooling from 400 to 100\,$^\circ$C (roughly 20 minutes), followed by the final annealing step (without sputtering) at 900\,$^\circ$C.  We can only speculate about the reason for the different outcome of sputtering at RT or slightly higher temperatures:  Any structural of morphological differences between samples sputtered at RT or slightly elevated temperatures should disappear after annealing at 900\,$^\circ$C.  Thus, differences of surface composition must be responsible for the domain boundaries seen after sputtering at RT.  It is well known that preferential sputtering of alloys leads to a modification of the chemical composition, and the composition can depend on the sputtering temperature: At elevated temperatures, the preferentially sputtered element can be replenished at the surface by diffusion, causing depletion of that element over a thicker region of the sample.  In addition, preferential sputtering can depend on the roughness of the sample, which also depends on the temperature. Assuming that Pt is preferentially found at steps,\cite{woudaPtRh} Pt would be sputtered preferentially on rougher samples with many steps, again modifying the surface composition depending on the sample temperature during sputtering. Unfortunately, the differences of surface composition are too small to be detected by AES.

Recently, a phase diagram was constructed from thermodynamical data of the Pt-Zr system.\cite{Gao} It shows that the Pt$_3$Zr phase has a narrow stability range ($\approx 0.5$\,\% at 1000\,K, even less at lower temperatures), i.e., phase separation occurs at very small deviations from the Pt$_3$Zr stoichiometry.  This supports our notion that the sample is sensitive to tiny variations of the stoichiometry.  Slight deviations from the ideal stoichiometry do not lead to nucleation of a completely different phase; this would be easily detected by different terrace heights and different atomic structure in STM.  Instead, we observe domain boundaries, which are probably enriched in one element.


\subsection{DFT study of the Pt$_3$Zr(0001) surface \label{secAlloyDFT}}

In order to benchmark the description of the substrate by DFT, we determined the optimal lattice parameters of the Pt$_3$Zr bulk using both the GGA (PBE) and vdW-DF (optB88) functionals. 
Table \ref{tab:Pt3zrbulk} shows the results of a fit to a Birch-Murnaghan equation of state, together with the experimental values from Ref.\ \onlinecite{PtZrLB}.
The comparison shows that the predicted geometries of the GGA (PBE) and vdW-DF (optB88) are rather similar; both slightly overestimate the experimental lattice parameters, which results in an overestimation of the volume by about 5.5 and 6.1\%, respectively.

\begin{table}[tb]\begin{center}
\begin{tabular*}{\columnwidth}{@{\extracolsep{\fill}} l r r r }
  \hline \hline
   & PBE & optB88 & exp. \\
  \hline
  $a$ (pm) & 572.9 & 574.2 & 562.4 \\
  $c$ (pm) & 936.4 & 938.0 & 921.3 \\
  Volume (nm$^{3}$)  & 0.2663 & 0.2678 & 0.2524 \\
  \hline \hline
\end{tabular*} \caption{Calculated and experimental lattice parameters and equilibrium  volume for bulk Pt$_3$Zr}\end{center}\label{tab:Pt3zrbulk}\end{table}


From the calculated surface energies of the two different terminations of Pt$_3$Zr(0001) we can determine whether the terminating layer should be of A type (fcc) or B/C type (hcp; see Fig.\ \ref{FigNi3Ti}).  To address this question, we compared the calculated surface energies for two differently terminated 5-layer slabs of Pt$_3$Zr by DFT (PBE functional). The calculations yield a surface energy of 0.73 eV per surface atom for the A-terminated slab (ABACA stacking, as shown in Fig.\ \ref{FigNi3Ti}). This is 66\,meV lower than for the C-terminated (CABAC stacked) slab. The relative arrangement of the first two layers is the same for both slabs; thus, the 3rd monolayer significantly influences the surface energy, stabilizing the A termination. The DFT result is in agreement with the experimental observation of an A-terminated surface.

The A-terminated Pt$_3$Zr surface (7-layer slab) shows a small rumpling in the top layers. The  surface Zr atoms are 15\,pm lower than the Pt. In the second monolayer, the Zr is 15\,pm higher than the Pt. The distance between the first and the second monolayer is 229.5\,pm, and between the second and the third monolayer 236.1\,pm (we define the interlayer distances as the difference of the average height of the atoms in each layer). Compared to the calculated bulk value of 234.1\,pm,  this corresponds to a 1.9\% contraction of the first, and a 0.4\% expansion of the second interlayer distance. The overall inward relaxation of the first monolayer, followed by a slight outwards relaxation of the 2nd ML is typical for metal surfaces. It is unusual, however, that the larger atom (Zr) in the surface buckles inwards. 
Nevertheless, the relative relaxations of Pt ($-0.4$\,\%) and Zr ($-6.7$\,\% with respect to the bulk interlayer distance of the alloy) in the Pt$_3$Zr(0001) surface are comparable to the pure metals [+1\,\% and $-6.4$\,\% for Pt(111) \cite{Zolyomi} and Zr(0001) \cite{Kwon}, respectively].

We also simulated a surface terminated by a full Pt layer, as observed experimentally after oxidation of the sample (see below). For constructing these Pt-terminated slabs we simply replaced zirconium in the surface layer by platinum. Again we find a higher stability for the A termination; albeit with a low surface energy difference of 6\,meV per atom as compared to the B/C termination. 
Also the Pt-terminated slab shows a slight buckling of the surface: The Pt atoms with a Zr neighbor in the 2nd ML are 7\,pm higher than the other Pt surface atoms. 
Similar to the A-terminated stoichiometric Pt$_3$Zr, the Zr atom of the second monolayer is situated 12\,pm closer to the surface than the Pt atoms of the same layer.  The first two interlayer distances in the Pt-terminated slab are 229.2\,pm and 233.1\,pm, i.e., both are contracted with respect to the bulk (by 0.4\% and 2.1\%, respectively).

\section{Z\lowercase{r}O$_2$ films on the P\lowercase{t}$_3$Z\lowercase{r}(0001)}\label{secOxide}

\subsection{Preparation and experimental investigations}\label{secOxideSTM}

\begin{figure}[tb]
  \centering
    {\includegraphics [width=7cm]
    {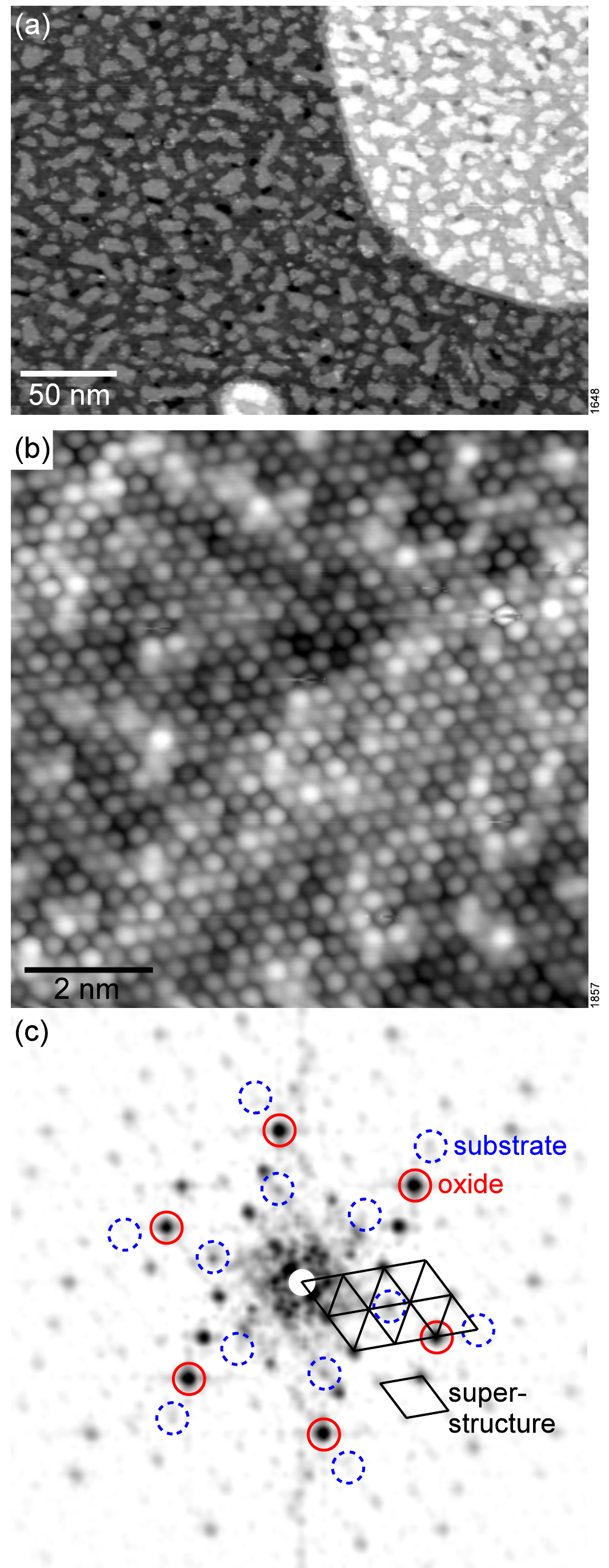} }
    \caption{(Color online) (a) STM image of the Pt$_3$Zr(0001) surface oxidized at 400\,$^\circ$C. (b) Atomically resolved image ($-2.5$\,mV, 0.45\,nA) of the oxide annealed at 1000\,$^\circ$C. (c) Fourier transform of (b).}
  \label{FigOxideSTM}
  \end{figure}

In order to prepare an ultrathin zirconia film at the alloy surface, we oxidized the sample at $\approx 10^{-7}$ mbar O$_2$ at 400\,$^\circ$C for 30 minutes. This treatment results in many islands, about 0.2\,nm in height, and a few holes of $\approx 0.2$\,nm depth [Fig.\ \ref{FigOxideSTM}(a)].   About 30--40\% of the surface is covered with islands; the total area of the holes is low ($\approx 1\%$).  A flat oxide-covered surface can be obtained by post-annealing at temperatures $\geq 750\,^\circ$C. If the annealing temperature is too high, $\geq 900\,^\circ$C, the oxide film becomes discontinuous, with a small fraction of uncovered metal; presumably oxygen diffuses into the bulk.  This is the only difference between films annealed at different temperatures above 750\,$^\circ$C.  STM images with atomic resolution are obtained at low tunneling voltages typical for ultrathin oxide films (cf.\ Ref.\ \onlinecite{KresseScience}), and the oxide shows good order with a lattice constant of $0.350 \pm 0.002$\,nm [Fig.\ \ref{FigOxideSTM}(b)].  The atoms imaged by STM show significant height differences of up to 50\,pm; almost every atom has at least one neighbor differing in apparent height by $\geq 20$\,pm.

Fig.\ \ref{FigOxideSTM}(c) shows the Fourier transform of  Fig.\ \ref{FigOxideSTM}(b). The spots of the oxide grid are marked by full circles. The outer broken circles correspond to the metal lattice of the substrate ($\approx 0.28$\,nm), and the inner broken circles to the unit cell of the ordered alloy ($\approx 0.56$\,nm). Not all of these substrate spots are apparent in the Fourier transform, but their positions can be ascertained by comparison with images of the substrate. The superstructure cell (1.2\,nm in real space) is marked by black lines. It is commensurate, each Fourier spot of the 0.28\,nm substrate lattice and the oxide coincides with the superstructure reciprocal lattice. In terms of the reciprocal lattice of the superstructure, the reciprocal lattices of the oxide and metal have side lengths of  $2\sqrt{3}$ and $\sqrt{19}$, respectively. Translated to real space, this means that the superstructure has a $(\sqrt{19} \times \sqrt{19})$R$23.4^\circ$ cell with respect to the 0.28\,nm metal atomic lattice (i.e., not taking into account the chemical order of the alloy), and the lattice constant of the oxide is $\sqrt{19}/(2\sqrt{3}) = 1.258$ times that of the metal. The rotation angle between the oxide and the substrate lattice is $30 - 23.4 = 6.6^\circ$. 
The spots of the 0.56\,nm alloy lattice appear at half-integer positions with respect to the reciprocal cell of the $(\sqrt{19} \times \sqrt{19})$R$23.4^\circ$ superstructure (which was referred to the 0.28\,nm lattice).  Thus, when including the chemical order of the alloy, the superstructure cell is twice as large in real space (2.4\,nm).
We have analyzed several STM images of different preparations and always found the same relation between substrate and oxide lattice.

The hexagonal lattice suggests formation of a ZrO$_2$(111) layer, as already found for films grown by deposition and oxidation of Zr on Pt(111)\cite{Maurice, Meinel03,Meinel04,Meinel06}. The lattice constant of 0.350\,nm found by STM is only 3\% lower than the interatomic distance in (111) layers of cubic ZrO$_2$ \cite{ZrO2Note}.  The charge-neutral building block of ZrO$_2$(111) is an O-Zr-O trilayer. From STM, we cannot determine the thickness of the oxide, i.e., whether one or more such trilayers are present.

To this end we compare the oxygen Auger signal of the ZrO$_2$ films with that of a RhO$_2$/Rh(111) trilayer oxide, which has a well-known structure with an in-plane lattice constant of 0.310\,nm as determined by surface x-ray diffraction.\cite{GustafsonRh111}  In contrast to the RhO$_2$ trilayer oxide, AES measurements of the ultrathin zirconia films are hampered by strong electron-stimulated desorption (ESD) of oxygen; we observe a 10\% decrease of the oxygen signal within 50\,seconds.  Correcting for ESD by extrapolating the oxygen signal back to the start of the measurement we obtain a ratio between the oxygen signals of the zirconia film and the RhO$_2$ trilayer of $0.63 \pm 0.1$.  This is slightly less than expected on the basis of the areal densities of the two oxides, $(a_{ZrO_2} / a_{RhO_2})^{-2} = 0.78$.  The discrepancy might be related to a shoulder observed at the high-energy side of the oxygen peak of RhO$_2$, which amounts to 15\% of the total O signal; the shoulder is possibly due to oxygen dissolved in the Rh bulk.  Nevertheless, these AES results clearly rule out the possibility of a ZrO$_2$ film thicker than a single trilayer. As our STM images always show the same superstructure and appearance of the oxide we conclude that the zirconia film always consists of a single trilayer.

During growth of the ultrathin zirconia film, Zr is liberated from the alloy.  Defining 1\,ML (monolayer) as the density of atoms in the alloy surface, each layer of the alloy contains 0.25\,ML Zr. To form the oxide, we need $(a_{ZrO_2} / a_{metal})^{-2} = 0.64$\, ML Zr.  Counting all metal atoms irrespective of their identity, this means that 64\% of the upper monolayer is removed. The remaining 36\% of the surface layer are present as islands unless mobility is high enough for mass transport to or from steps [cf.\ Fig.\ \ref{FigOxideSTM}(a)].  Mass transport of Zr into the oxide layer also results in a Pt-enriched near-surface region of the alloy.  The narrow range for the Pt$_3$Zr structure in the phase diagram (mentioned above, cf.\ Ref.\ \onlinecite{Gao}) suggests that replenishing the missing Zr needs long diffusion paths to keep the alloy composition within the Pt$_3$Zr stability range.  Such long-distance diffusion will need extremely high temperatures, otherwise the alloy surface remains Pt rich.  The amount of Zr needed for creation of the oxide layer corresponds to the Zr content within $0.64/0.25 \approx 2.5$\,ML of the alloy.

\begin{figure}[tb]
  \centering
    {\includegraphics [width=8cm]
    {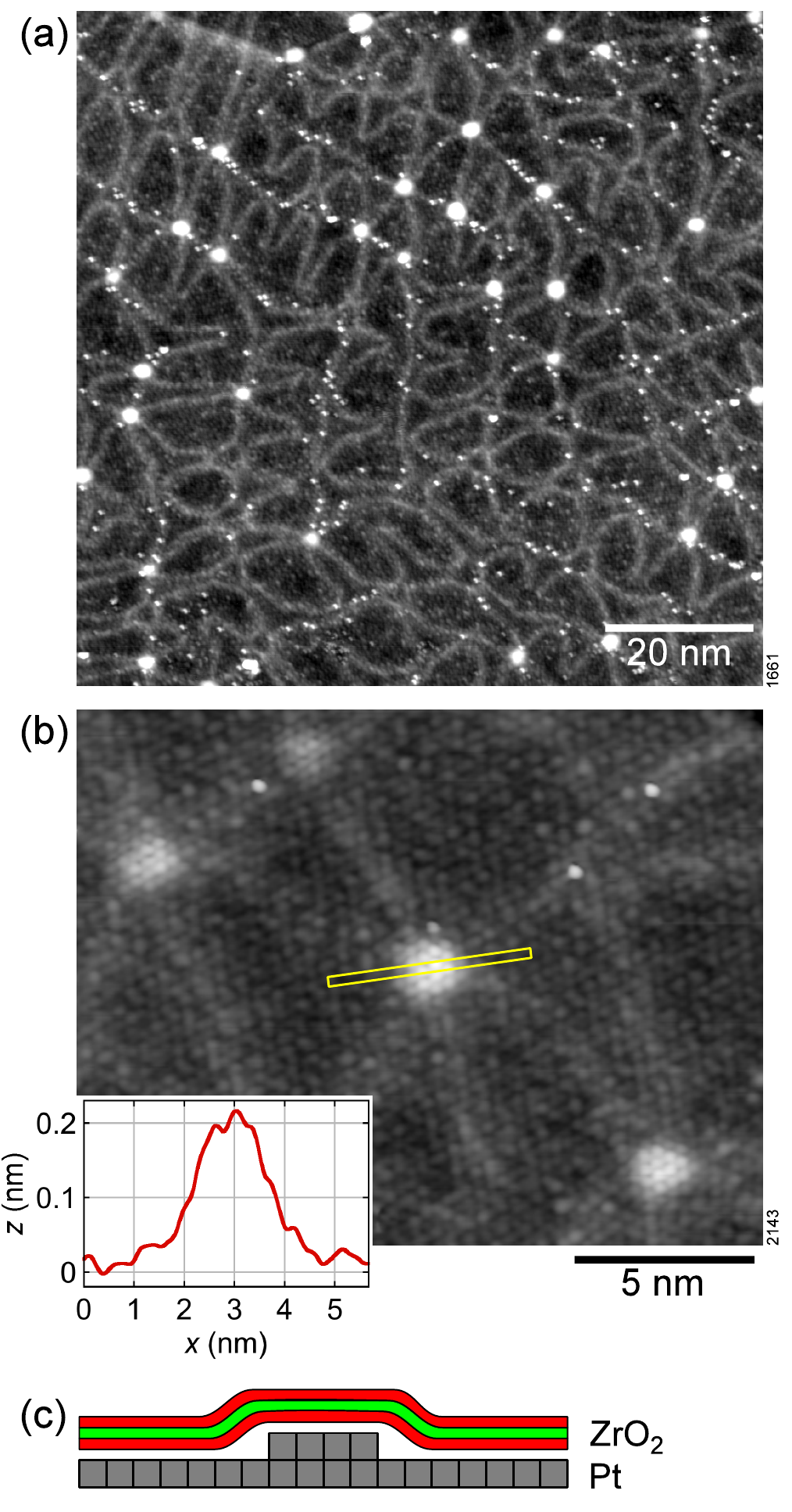} }
    \caption{(Color online) (a), (b) STM images of the Pt$_3$Zr(0001) surface oxidized at 400\,$^\circ$C and annealed at $T \ge 850\,^\circ$C [(a): 2\,V, 0.1\,nA; (b): $-14$\,mV, 0.5\,nA]. The inset shows a line profile taken across the protrusion in the middle of the image.}
  \label{FigPtRec}
  \end{figure} 

The presence of at least one Pt layer below the oxide also explains the pattern of weak ridges  on the surface [Fig.\ \ref{FigPtRec}(a)].  Pt(111) is know to be on the verge of reconstruction,\cite{MichelyPt111rec} where the first ML contracts in a manner similar to Au(111). With an interatomic distance of 0.281\,nm, the alloy lattice is larger than that of pure Pt (0.2775\,nm), so it is natural that the first Pt layer contracts with respect to the underlying bulk lattice.  As is usual for structures consisting of a first layer with a lattice constant smaller than the underlying bulk, the bright ridges mark the positions where the first-layer (Pt) atoms are located in bridge sites of underlying substrate lattice. The alternating wide and narrow dark regions between the ridges indicate that one of the two stacking sequences (Pt in hcp or fcc hollow sites) is preferred.

Our STM images also show small $\approx 0.2$\,nm high protrusions in the oxide lattice [Fig.\ \ref{FigPtRec}(b), see also the linescan in the inset].  These protrusions are always found at the nodes of the Pt dislocation network.  The height of these protrusions is close to that of a metal layer, suggesting that these protrusions are caused by a small metal island below the oxide.  In contrast to normal substrate steps, where we see a rather sharp boundary, the oxide lattice runs smoothly over these protrusions like a carpet [Fig.\ \ref{FigPtRec}(c)].  This observation and the fact that such defects should quickly disappear at the annealing temperature indicates that these are metal islands resulting from expulsion of metal atoms from the surface while cooling to room temperature.  Possibly this is caused by a lower coefficient of thermal expansion of the dense (thus hard), reconstructed Pt(111) layer compared to the bulk of the alloy.  The junctions of the dislocation lines mark points where dislocations reach the surface; these sites should be nucleation sites for any metal atoms diffusing beneath the oxide carpet at high temperature, and, in addition, the more weakly bound metal atoms at these defect sites should be also the first ones to get expelled when the Pt layer contracts.

We further observe a few bright atom-sized defects in the oxide [e.g., two near the top right corner of Fig.\ \ref{FigPtRec}(b)].  These defects also show a clear correlation with the ridges (dislocations) of the Pt reconstruction.  The nature of these point defects is unclear.

Fourier transforms [Fig.\ \ref{FigOxideSTM}(c)] show spots in the positions corresponding to the 0.56\,nm lattice of the ordered alloy; in some STM images (especially when taken at positive sample bias) this is even the dominating Fourier component.  For a thick Pt layer on top of the ordered alloy the 0.56\,nm modulation should be smoothed out; thus, the Pt layer is probably only one ML thick, and the second substrate layer is already an ordered alloy.

\begin{figure}[tb]
  \centering
    {\includegraphics [width=8cm]
    {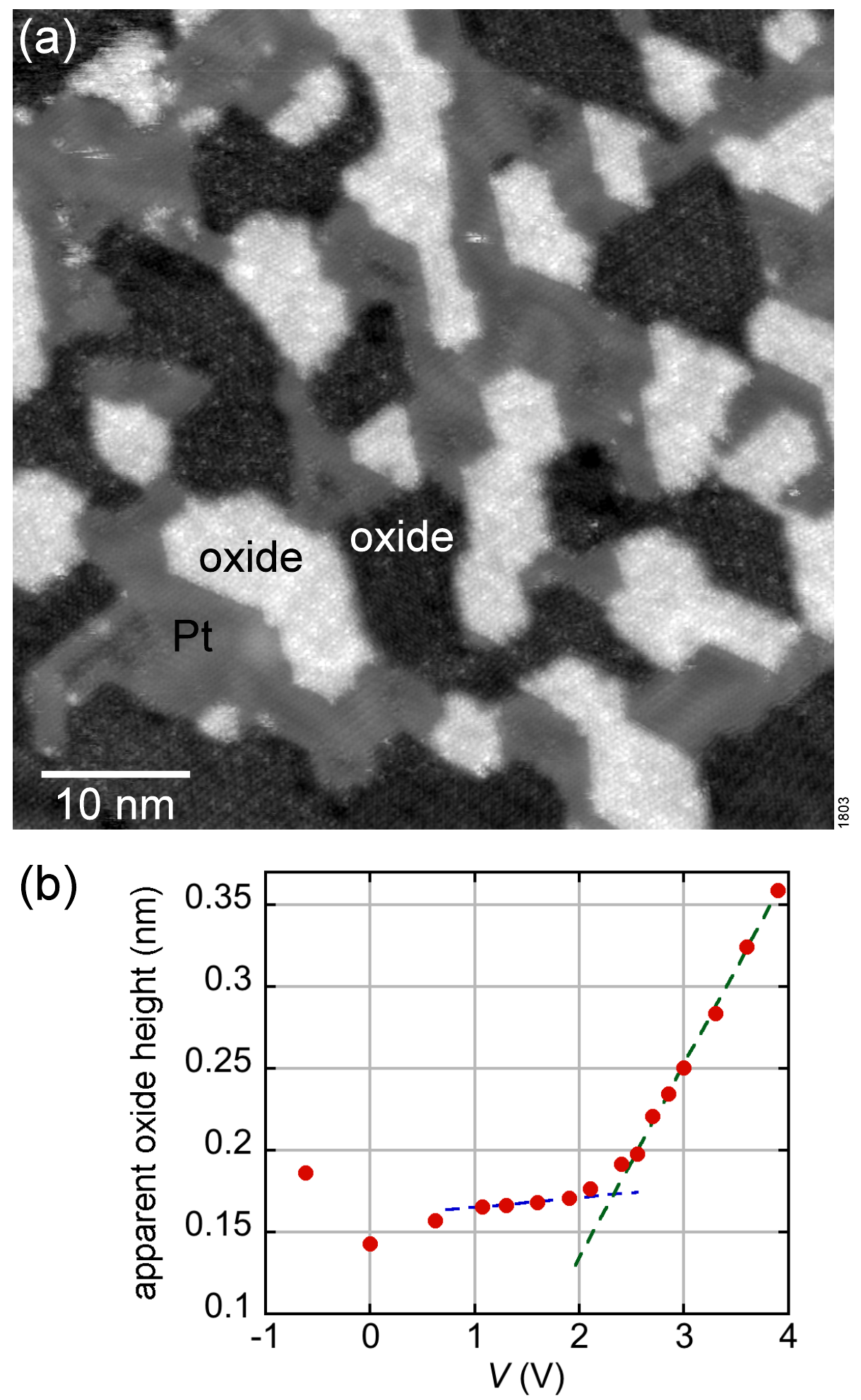} }
    \caption{(Color online) (a) STM image of a surface partially covered by ZrO$_2$ ($-2$\,mV, 5.8\,nA). The height difference between the two oxide levels is $\approx 0.22$\,nm, corresponding to one metal layer. (b) Apparent height of an oxide island [bright region marked ``oxide'' in (a)] with respect to the metal (marked ``Pt''). The data were taken with a tunneling current of 0.14\,nA, except for the data point at $-2$\,mV, 5.8\,nA. The broken lines are guides to the eye. }
  \label{FigOxHeight}
  \end{figure} 

Shorter oxidation (5 instead of 30 minutes), and no post-annealing results in a surface partially covered by oxide.  The O/Pt AES peak ratio of this surface was roughly half of that of a fully-oxidized high-temperature annealed sample, and STM shows $\approx 35$\% bare metal surface [Fig.\ \ref{FigOxHeight}(a)].  It is interesting to note that the uncovered metal surface also shows the dislocation lines of the Pt reconstruction [only weakly visible in Fig.\ \ref{FigOxHeight}(a) because of their low corrugation compared to the step heights]. Thus, the first monolayer consists of Pt in the uncovered areas as well; the amount of Zr needed for formation of the incomplete oxide layer is still higher than that in one monolayer of the alloy.  The Pt surface also shows height modulations due to the 0.56\,nm lattice of the underlying alloy.

Fig.\ \ref{FigOxHeight}(a) shows two levels of oxide (bright and very dark gray); the height difference equals one metal interlayer distance (0.23\, nm).  This is a consequence of mass transport due to the different metal densities in the alloy and the oxide.

We now focus on the electronic properties of the oxide. A first impression of its metallic or insulating nature can be gained by studying the apparent height of the oxide on the metal as a function of tunneling voltage.  When tunneling into the band gap, an ultrathin insulator appears thinner than its geometric height. \cite{Bertrams,Hebenstreit}  When tunneling into states above the conduction band minimum, the apparent height strongly increases. \cite{Napetschnig}  The sample with partial oxide coverage [Fig.\ \ref{FigOxHeight}(a)] allows us to compare the apparent height of the oxide with that of the metal substrate (Pt terminated).
The apparent height of the oxide increases very slowly at low voltages; above $\approx 2.5$\,V the height rises sharply and almost linearly [Fig.\ \ref{FigOxHeight}(b)].  This behavior is very similar to the one observed for ultrathin alumina films on NiAl(110).\cite{Napetschnig}  From the intercept of the two broken lines in Fig.\ \ref{FigOxHeight}(b), we estimate that the conduction band minimum of the oxide is at $\approx 2.3$\,eV above the Fermi level.

\subsection{DFT study of the ultrathin Z\lowercase{r}O$_2$ films}\label{secOxideDFT}

%
%
%
%

We have first calculated the properties of a free-standing ZrO$_2$(111) trilayer. DFT gives an in-plane lattice constant of 330\,pm, 
9\,\% less than the in-plane distances in cubic ZrO$_2$(111) (DFT value: 363.2\,pm). The in-plane contraction is accompanied by a 31\%  expansion perpendicular to the plane, with calculated interlayer distances between the Zr and O layers of 97.4\,pm in the trilayer film, compared to 74.1\,pm in cubic bulk ZrO$_2$.

\begin{figure*}[tb]
  \centering
    { \includegraphics [width=17cm]{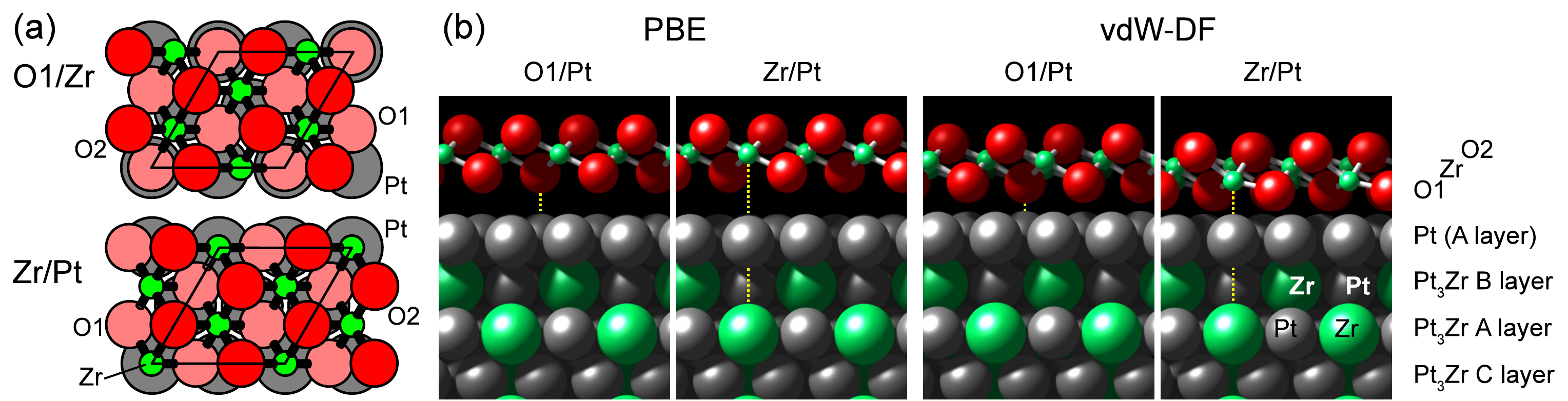} }
    \caption{(Color online) Calculated models for a ZrO$_2$(111) trilayer adsorbed on Pt/Pt$_3$Zr(0001): (a) Top view; the rhombus shows the unit cell. (b) Side view; vertical dotted lines indicate atoms on top of each other.}
  \label{DFTmodels}
  \end{figure*}

To decrease the computational effort involved in calculating the full $(\sqrt{19} \times \sqrt{19})$R$23.4^\circ$ superstructure of the ZrO$_2$/Pt$_3$Zr(0001) observed in experiment, we used a smaller model for the DFT calculations. The diagonal of the unit cell of the free-standing ZrO$_2$(111) (572\,pm) fits the DFT lattice constant of the alloy ($a = 572.9$\,pm), so we simply use one alloy unit cell with $(\sqrt{3} \times \sqrt{3})$ oxide cells on top (Fig.\ \ref{DFTmodels}).  By keeping roughly the lattice constants of the free-standing ZrO$_2$(111) trilayer, this approach overestimates the lateral contraction of the oxide film on the metal substrate; according to our STM data the oxide is contracted by only 3\%.  The essential properties of the system should nevertheless be reproduced well by the calculations.

We have compared two lateral adsorption configurations of the ZrO$_2$ film (Fig.\ \ref{DFTmodels}). Either one of the Zr atoms of the trilayer, or one of the the lower oxygen atoms (O1) is positioned on top of a substrate Pt atom.  For both configurations, the PBE adsorption energies are low and degenerate within our computational accuracy: 300\,meV (Zr above Pt) and 299\,meV (O1 above Pt) per ZrO$_2$ formula unit (Table \ref{DFTresultsOxide}). The calculated interlayer distances between the lower O layer (``O1'') of the ZrO$_2$ trilayer and the substrate are large, more than 300\,pm, again indicating a very weak adsorption.  Nevertheless, the interaction with the substrate induces a slight vertical contraction of the oxide film, with the Zr-O interlayer distances reduced by 2--3\,pm compared to the free-standing film, and a slight buckling of the Zr layer. The latter is stronger if a Zr atom sits directly above a Pt atom (configuration ``Zr/Pt'' in Table \ref{DFTresultsOxide}); this Zr atom moves down, towards the metal substrate. Within the computational accuracy, the buckling of the Pt substrate atoms is the same as in the bare substrate (7\,pm).

\begin{table}[tb]
\begin{center}

\begin{tabular*}{\columnwidth}{@{\extracolsep{\fill}}l rr rr}
\hline\hline
Functional & \multicolumn{2}{c}{PBE} & \multicolumn{2}{c}{vdW-DF} \\
Configuration  & O1/Pt & Zr/Pt & O1/Pt & Zr/Pt \\
  \hline
$d_\mathrm{O2-Zr}$ (pm)      & 95 & 95 & 96 & 96 \\
$d_\mathrm{Zr-O1}$ (pm)      & 95 & 95 & 95 & 94 \\
$d_\mathrm{O1-Pt}$ (pm)      & 349 & 337 & 266 & 244 \\
$b_\mathrm{Zr}$ (pm)         &  2 &  9 & 56 & 71 \\ 
$b_\mathrm{O1}$ (pm)         &  0 &  0 & 5  & 0 \\ 
$b_\mathrm{Pt}$ (pm)         &  8 &  7 & 7  & 1 \\
$E_\mathrm{ads}$ (meV)     & -299 & -300 & -338 & -385\\
\hline\hline
\end{tabular*} \caption{Calculated interlayer distances $d$, buckling $b$, and adsorption energies per ZrO$_2$ formula unit, for two configurations with O1 or Zr on top of Pt. Interlayer distances refer to the average height of the atoms in each layer, and buckling values are measured between the lowest and highest atom of each layer. The lower and upper oxygen layers in the oxide are named O1 and O2, respectively; and the buckling of the O2 layer is below 1\,pm.}
\end{center}
\label{DFTresultsOxide}
\end{table}

As standard DFT neglects non-local contributions, we have performed additional calculations using the van der Waals DFT (vdW-DF), which also gives an improved description for physisorbed overlayers \cite{MittendorferGraphene}. Indeed, the vdW-DF results in a significantly stronger adsorption. The adsorption energies show a moderate increase by 40--85\,meV (O1/Pt and Zr/Pt geometries).  More significant is the decrease of the distance to the substrate by 83--93\,pm, bringing the substrate-oxide distances closer to typical bond lengths.

The closer distance results in a significant buckling of the Zr layer.  In the Zr/Pt geometry, the Zr atom directly above a Pt substrate atom moves down by 47\,pm (compared to the center between the O1 and O2 layers).  This buckling results in a Pt-Zr distance of 293\,pm, which is  comparable to the Pt-Zr distance in the alloy ($a/2 = 286$\,pm with this functional).  In the O1/Pt geometry, the Zr atom in the oxide closest a Pt atom in the substrate buckles down by 35\,pm (Pt-Zr distance: 338\,pm).  This indicates that the oxide film binds to the substrate via the Zr atoms, not the oxygen, even though the O1 atoms reside much closer to the substrate than the Zr.  In spite of the large movement of the Zr atom closest to Pt, the force exerted by the Pt-Zr bond is small: The Pt atom directly below the Zr in the oxide moves up by only 5\,pm upon adsorption.  Thus, the ultrathin oxide is rather soft with respect to distortions of its Zr sublattice.  This type of distortion is different from the soft phonon mode leading to instability of cubic bulk ZrO$_2$ below $\approx 2500\,$K; there the oxygen atoms are displaced in an initially rigid Zr sublattice \cite{souvatzis}.

While one Zr atom in the ZrO$_2$ trilayer buckles down, the two others move up by roughly half that amount, keeping the overall O1-Zr and Zr-O2 interlayer distances almost unchanged.  This can be explained by in-plane displacements of the O atoms. When a Zr atom directly above a Pt moves down, its three lower oxygen neighbors remain at the same height above the substrate and move laterally away from this Zr to make space.  This increases the O-O distance in the triangle below the Zr from 331 to 357\,pm.  The oxygens above this Zr atom also remain in the same plane, but move closer together (O-O distances 312 pm).  Now the other two Zr atoms experience a compressed oxygen layer below and expanded O-O distances above; to keep reasonable Zr-O bond lengths they move up.

In the O1/Pt configuration (vdW-DF geometry), the distance between the on-top O1 atom and the Pt below is 258 pm.  This O atom moves down by 5\,pm and the Pt atom moves up by 5\,pm, indicating that the O-Pt interaction is also attractive, but the resulting O-Pt distance is still significantly larger than a typical O-Pt bond length (204\,pm in PtO; Ref.\ \onlinecite{McBride}).  Comparing the bonding energies and the substrate-oxide distances of the Zr/Pt and O1/Pt configurations also shows that the oxide binds much more strongly via its Zr atoms than via the oxygen.

\begin{figure}[tb]
  \centering
    {
    	\includegraphics [width=8cm]{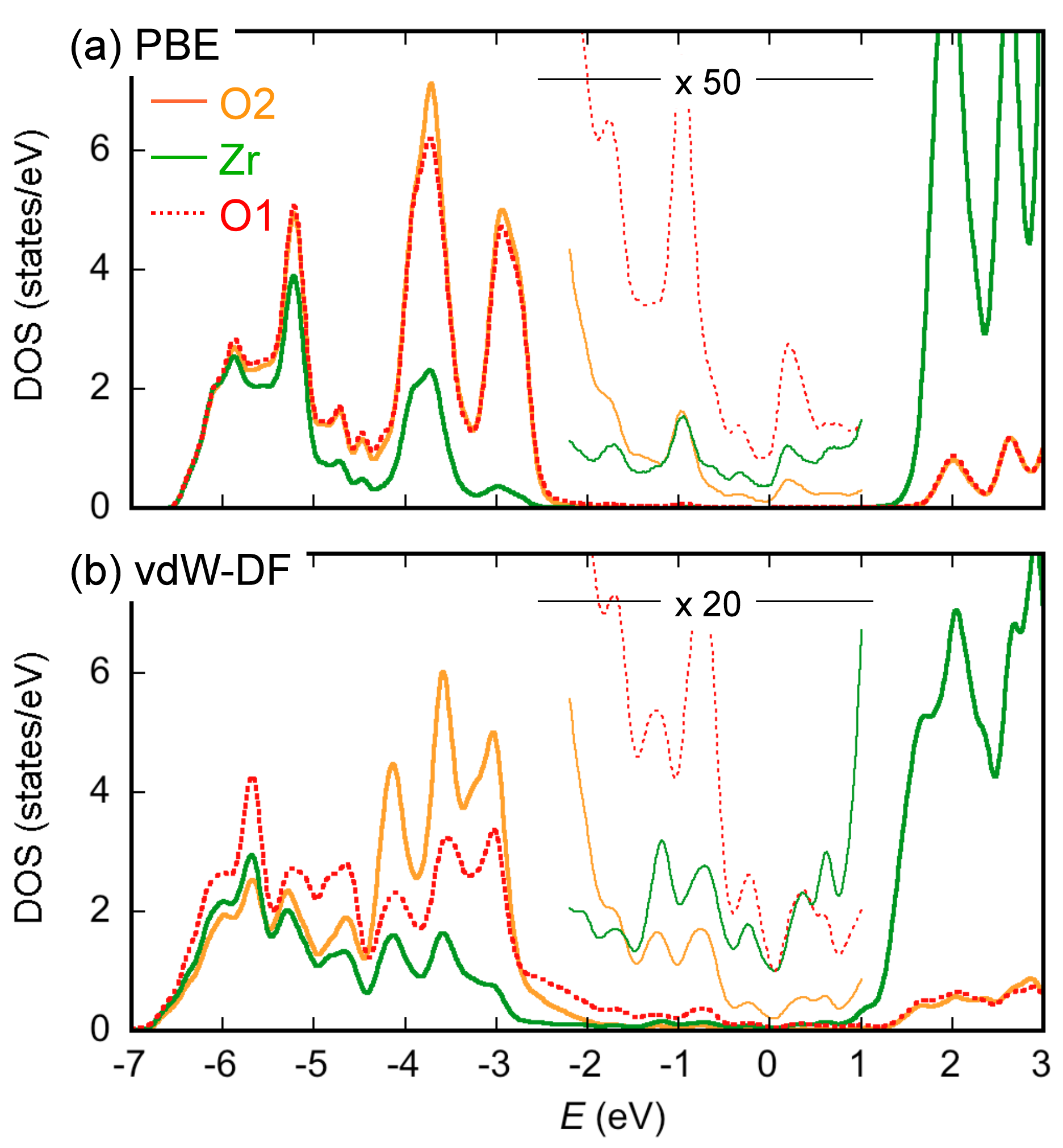}
    }
    \caption{(Color online) Projected density of states for ZrO$_2$ adsorbed on Pt/Pt$_3$Zr(0001), averaged over the O1/Pt and Zr/Pt configurations for (a) the PBE and (b) the vdW-DF geometry.  For both geometries, the DOS was evaluated with the PBE functional.}
  \label{DOS}
  \end{figure}

Figure \ref{DOS} gives an overview of the electronic density of states (DOS) of the ZrO$_2$ film for both, the PBE and vdW-DF configurations.  The experimentally observed large $(\sqrt{19} \times \sqrt{19})\mathrm{R}23.4^\circ$ cell contains many different lateral positions of the oxide on the substrate, we thus averaged the DOS over the O1/Pt and Zr/Pt simulation cells for each configuration.  In both, the PBE and vdW-DF configurations, we find a band gap of about 4 eV for the trilayer oxide.  The DOS is essentially zero in the vicinity of $E_F$. Minor contributions in the band gap arise from hybridization with the metal support.  As expected, these contributions are strongest for the lower oxygen atoms (O1, dotted), and increase with decreasing distance to the substrate (from the PBE to the vdW-DF configuration).  The DFT calculations predict a conduction band minimum at $\approx 1.5$\,eV.  When comparing this result with the experimental value ($\approx E_F + 2.3$\,eV; Fig.\, \ref{FigOxHeight}), it should be noted that DFT systematically underestimates band gaps, thus the experimental and calculated values can be considered in reasonably good agreement.

We have also simulated STM images for both, the PBE and the vdW-DF geometry, by plotting the local density of states (LDOS) above the surface between $E_F$ and $E_F + 0.5$\,eV.  We find that the highest LDOS is found above the upper atoms (O2) in the PBE geometry, but above the Zr atoms in the vdW-DF geometry.  This result is directly related to the DOS near $E_F$ plotted in Fig.\ \ref{DOS}, where the Zr DOS increases more strongly with decreasing oxide-substrate distance than the O2 DOS (compare the PBE and vdW-DF geometries).  The low DOS in the bandgap and the low tunneling resistance used for atomic resolution [Fig.\ \ref{FigOxideSTM}(b)] imply a short distance between the tip and the oxide.  Together with the soft Zr sublattice mentioned above, we consider it likely that the experimental STM images are dominated by tip-sample interaction.  Thus, the LDOS is only of limited value for predicting whether STM shows the oxygen or the Zr atoms as protrusions.

Nevertheless, the STM results help us to decide whether the PBE or vdW-DF functional better describes this system:  The large buckling of the oxide seen by STM cannot be reconciled with the PBE results, which predict flat layers for oxygen and Zr.  Thus we consider the vdW-DF geometry a more realistic description of the system than PBE.  The large buckling observed by STM also indicates that STM shows the Zr atoms as protrusions.

\section{Conclusions}

We have characterized the Pt$_3$Zr(0001) surface by STM and DFT and grown ultrathin ZrO$_2$ films by oxidation of this alloy.  The film consists of a O-Zr-O trilayer. Oxidation is self-limiting, similar to ultrathin alumina films obtained by oxidation of NiAl alloys, indicating that O$_2$ cannot dissociate on the trilayer oxide.  Incorporation of Zr into the oxide leads to a Pt layer at the interface, which reconstructs by contracting.  DFT calculations show that the oxide binds to the substrate via some of its Zr atoms, causing a large rumpling of the Zr layer in the oxide. The O layers remain rather flat and the average height of the Zr layer remains centered between the two O layers.  Strong oxide-support bonding via the metal cations is very different from the surface oxides of the late transition metals.  All these surface oxides bind to the surface via oxygen,\cite{LundgrenJPCM} even when the oxygen-metal interlayer distances are much closer than in the ZrO$_2$ film, putting the metal atoms close to the substrate.\cite{LundgrenPd111}.

Our study also indicates that DFT-only calculations of an oxide-metal interface can give misleading results; non-local contributions (van der Waals corrections) play an important role and must be considered for obtaining the correct result.  Using vdW-DF (optB88 potential) leads to a significant change of adsorption height, from physisorption to bonding distances more typical for chemical bonds, but not to a dramatic increase of the adsorption energy (when compared to PBE calculations).  This is similar to recent calculations for graphene/Ni(111).\cite{MittendorferGraphene}.  Finally, the new and simple approach towards preparation of ultrathin zirconia films presented in this work should pave the way to detailed atomic-scale studies of the surface properties of this important material.

\section{Acknowledgements}
This work was supported by the Austrian Science Fund (FWF) under project number F45.
The Vienna Scientific Cluster (VSC) is acknowledged for computer time.

\begin{thebibliography}{99}
\bibitem{ZrO2Note}
This value is based on a ZrO$_2$ lattice constant of $a_\mathrm{bulk}=0.509$\,nm, Ref.\ \onlinecite{KatzZrO2}; the contraction would be 4\% when comparing to Ref.\ \onlinecite{PlocZrO2} ($a_\mathrm{bulk}=0.5135$\,nm).
\end{thebibliography}

\end{document}